# Elimination of HIV in South Africa through expanded access to antiretroviral therapy: Cautions, caveats and the importance of parsimony

Brian G. Williams

South African Centre for Epidemiological Modelling and Analysis (SACEMA), Stellenbosch, South Africa
Correspondence to BrianGerardWilliams@gmail.com

## Abstract

In a recent article Hontelez et al.[1] investigate the prospects for elimination of HIV in South Africa through expanded access to antiretroviral therapy (ART) using STDSIM, a micro-simulation model.[2] One of the first published models to suggest that expanded access to ART could lead to the elimination of HIV, referred to by the authors as 'the Granich Model', was developed and implemented by the present author.[3] The notion that expanded access to ART could lead to the end of the AIDS epidemic gave rise to considerable interest and debate and remains contentious. In considering this notion Hontelez et al.[1] start by stripping down STDSIM to a simple model that is equivalent to the model developed by the present author[3] but is a stochastic event driven model. Hontelez et al.[1] then reintroduce levels of complexity to explore ways in which the model structure affects the results. In contrast to our earlier conclusions[3] Hontelez et al.[1] conclude that universal voluntary counselling and testing with immediate ART at 90% coverage should result in the elimination of HIV but would take three times longer than predicted by the model developed by the present author.[3] Hontelez et al.[1] suggest that the current scale-up of ART at CD4 cell counts less than 350 cells/μL will lead to elimination of HIV in 30 years. I disagree with both claims and believe that their more complex models rely on unwarranted and unsubstantiated assumptions.

## Analysis

Hontelez et al.[1] take an interesting and novel approach to examining the extent to which model structure affects the predictions of a model. The start from a very detailed micro-simulation model, strip it down so that it resembles a simple compartmental model, fit this model to the available data and then reintroduce various levels of complexity until they recover the full micro-simulation model. However, their analysis appears to be flawed in important ways and these flaws need to be corrected.

In their Model A,[1] which most closely resembles my model[3] the steady state prevalence without ART is 15% and the incidence is 2% *p.a.* compared to 15% and 1.4% *p.a.*, respectively, for my deterministic model[3] (panel A in Figure 1). Since, prevalence is approximately equal to incidence times the duration of disease of 10 years the authors appear to have overestimated the incidence. This is a minor consideration and scaling their incidence down by 35% gives good agreement with my deterministic model[3] before 2005 and after 2013 (panel B in Figure 1). Between 2005 and 2013 their Model A differs significantly from my model (panel B in Figure 1) probably because their Model A fails to include the current scale up of ART which has now reached about 30% of all prevalent cases.[4] If Hontelez et al. were to make realistic assumptions in this regard the two models would be in close agreement.

The Hontelez et al. Model B[1] includes age-structure and heterogeneity in transmission by stage of infection. This increases the time to elimination most probably because they assume that in the absence of ART about 16% of transmission takes place during early acute/early infection which lasts for 3 months. However, available virological data suggest that acute infection lasts for not more than two weeks and there is no convincing evidence that acute/early infection contributes significantly to overall transmission[5] so that this assumption should be revised.

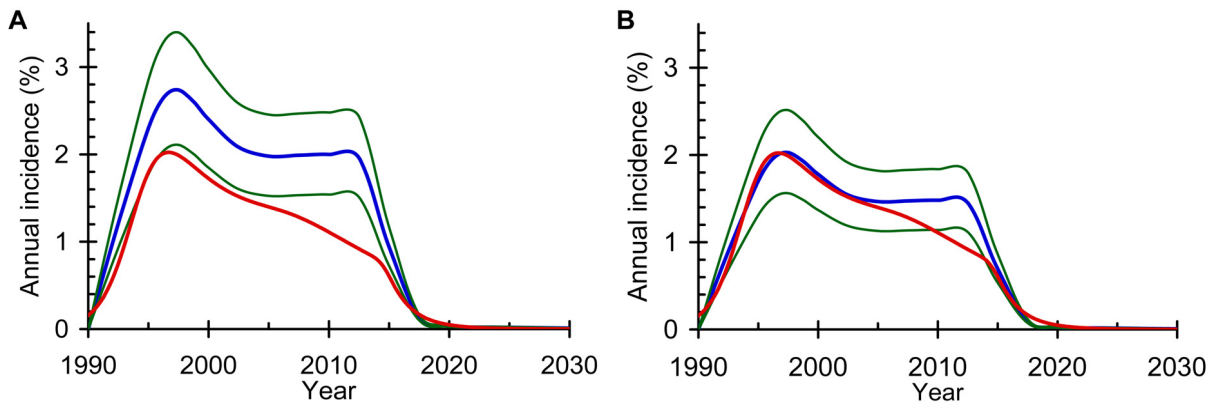

Figure 1. The estimated incidence of HIV in South African adults. A. Blue line: Model A of Hontelez et al.[1]; Green lines: their confidence limits; Red line using my deterministic model.[3] B. As in A with the incidence from the Hontelez model scaled down by 35%.

The Hontelez et al.[1] Model C explicitly introduces heterogeneity in sexual behaviour, male circumcision coverage, the effect of sexually transmitted infections that act as cofactors for HIV transmission, and increases in condom use; their model D introduces CD4 cell count decline with disease progression and ART coverage at the current level of scale up. However, their models C and D suggest that the incidence of infection fell by 30% between January 1999 and July 1999, a remarkable result. On closer inspection this appears to follow from the assumption (their Figure S4, Model C[1]), that condom use increased from 0% in January 1999 to 22% six months later, after which it remained unchanged. To support this claim they cite two studies. The first[6] shows a steady increase in the proportion of men and women who reported using a condom at last sex from 31% in 2002 to 65% in 2008; the second[7] shows an increase in condom use among women from 15% in 1998 to 70% in 2009. Neither is consistent with their assumptions. One may speculate that they are driven to make this assumption because the authors Models C and D rely on unrealistic assumptions about heterogeneity in sexual behaviour and, to stop the prevalence from continuing to rise after the year 2000, they make an assumption about the increase in condom use which makes little sense. It would be interesting to see the effect on their model predictions with realistic assumptions about changes in condom use over time.

In spite of their unusual and somewhat surprising assumptions it is encouraging to see that their model A,[1] in which the assumptions are soundly based, supports the earlier conclusions of our model that regular testing and early treatment could lead to the elimination of HIV provided one can achieve high rates of testing and high levels of adherence. It would be interesting to see how their more detailed models change under realistic assumptions, especially in relation to the impact of condom use.

The challenge now is to mobilize the political will and the financial support to make early treatment available to all that want it in order to save lives, save money and stop AIDS. Waiting to eliminate HIV when we already have the scientific evidence and tools to do so is not sound public health policy and will result in many unnecessary deaths.